\documentclass[aps,twocolumn,preprintnumbers,amsmath,amssymb]{revtex4}

\usepackage{graphicx}
\usepackage{dcolumn} 
\usepackage{bm}      
\usepackage{url}
\usepackage{amsmath}
\begin{document}

\title{Threshold Error Penalty for Fault Tolerant Quantum Computation with Nearest Neighbour Communication}

\author{Thomas~Szkopek,
        P.~Oscar~Boykin,
        Heng~Fan,
        Vwani~P.~Roychowdhury
        Eli~Yablonovitch,
        Geoffrey~Simms,
        Mark~Gyure,
        and~Bryan~Fong}
\thanks{T. Szkopek, H. Fan, V.P. Roychowdhury and E. Yablonovitch are with University of California, Los Angeles.}%
\thanks{P.O. Boykin is with University of Florida, Gainesville.}
\thanks{G.Simms, M. Gyure and B. Fong are with HRL Laboratories}

\begin{abstract}
The error threshold for fault tolerant quantum computation with
concatenated encoding of qubits is penalized by internal
communication overhead. Many quantum computation proposals rely on
nearest-neighbour communication, which requires excess gate
operations. For a qubit stripe with a width of $L+1$ physical qubits
implementing $L$ levels of concatenation, we find that the error
threshold of $2.1\times 10^{-5}$ without any communication burden is
reduced to $1.2\times 10^{-7}$ when gate errors are the dominant
source of error. This $\sim 175$X penalty in error threshold
translates to an $\sim 13$X penalty in the amplitude and timing of
gate operation control pulses.
\end{abstract}

\maketitle

\section{Introduction}

A critical architectural issue for quantum computation is the
internal communication of quantum information within the processor.
There are a variety of proposed quantum processor implementations
with different mechanisms for internal communication. For instance,
the linear ion trap proposal of Cirac and Zoller \cite{cirac95}
involves physical motion of massive ions for internal communication,
as do proposals using more complex ion trap structures
\cite{kielpinski02}. Alternative proposals involve using photons and
cavity QED for communication \cite{cirac97}. The cavity QED approach
has been extended to the solid state \cite{imamoglu99, wallraff04}.
Even direct transport of information carrying electrons has been
suggested for the solid state \cite{burkard00,recher00}.

Our paper is motivated by another class of quantum computation proposals that rely upon local
communication through nearest neighbour interactions \cite{loss98,kane98,vrijen00}. For instance,
communication among electron spins in semiconductors can be performed with sequential SWAP gate
operations, generated by a controlled Heisenberg exchange between adjacent electrons. An appealing
feature of the SWAP operation is that it is generated by the very same two-qubit interaction used
for computational operations. Also, a substantial degree of parallelism can be employed. However,
the protection of qubits with concatenated error correction requires communication between a number
of physical qubits that grows exponentially with concatenation level. This exponential increase in
SWAP operations might suggest that concatenated error correction will fail to reduce the logical
qubit error rate. Gottesman \cite{gottesman00}, and Aharanov and Ben-Or \cite{aharanov99} have
pointed out that a threshold error exists despite an exponential increase in logical gate count
with concatenation level $L$, although no attempt was made to quantify what that threshold might
be. In this paper, we estimate that threshold.

The main result we report here is that the number of nearest
neighbour communication operations is merely a constant factor over
and above the necessary logical operations for error correction at
each concatenation level $L$. Our estimated error thresholds are
summarized in Table \ref{table:errors}. We analyzed in detail
fault-tolerant error correction with a concatenated 7-qubit
Calderbank-Shor-Steane (CSS) code \cite{calderbank96,steane96} on a
linear qubit stripe with a width of $L+1$ physical qubits for $L$
levels of concatenation, and find an $\sim 175$ fold reduction in
threshold gate operation error due to nearest neighbour
communication overhead. This translates to an $\sim 13$ fold
increase in accuracy of control pulse amplitude and timing in gate
operations. Although nearest neighbour communication incurs a
significant penalty in the requisite experimental accuracy of qubit
gate operations, it is not a fundamental obstacle to fault-tolerant
computation in the solid-state. Our analysis is in general agreement
with the recent work of Svore \textit{et al.} \cite{svore04}, who
also show that internal communication with local interactions incurs
an error threshold penalty, although they do not fully account for
all communication steps.

Our paper is organized as follows. In the first section, we describe the underlying architecture of
a quantum processor composed of electron spin qubits, including a description of the physical
layout of electron spin qubits and their grouping into concatenated CSS logical qubits. We describe
a fault-tolerant error correction protocol in the second section. Our protocol implements error
recovery without direct measurement. In the third section, we calculate the threshold error for
gate operations under our error correction protocol, with various assumptions about available
resources. The fourth section considers the relation between control pulse accuracy and gate error
thresholds.

\section{Layout Architecture}

\begin{figure}
\centering
\includegraphics[scale=0.5]{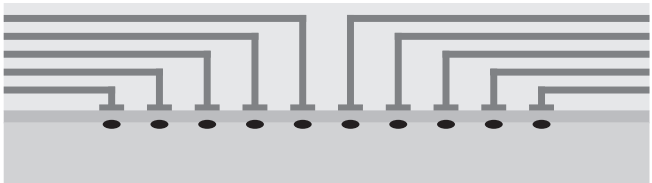}
\caption{\label{fig:qubitstripe} A schematic representation showing how the number of available
metal wire layers limits the width of a 2-D qubit array to only about 10-20 qubits. }
\end{figure}

\begin{figure}
\centering
\includegraphics[scale=0.5]{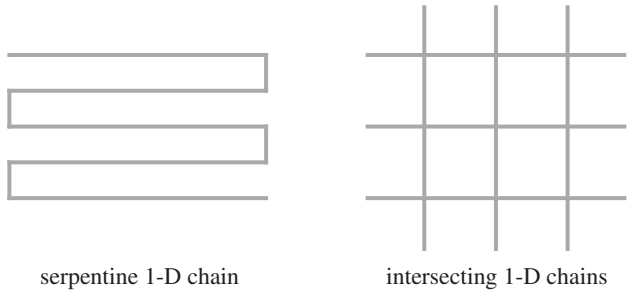}
\caption{\label{fig:serpentine} The requirement for gate electrode access to qubits restricts the
layout to stripes of either serpentine or intersecting geometry. }
\end{figure}

Given the problem of internal communication in a quantum processor, a higher dimensional
architecture is preferred because it would allow qubits to be as close as possible. However, there
must be access by control wires, thus limiting the packing geometry. Fig. \ref{fig:qubitstripe}
shows a schematic cross-section of a 2-D semiconductor qubit array controlled by gate electrodes
accessing qubits from the side. The number of vertical stacked control electrodes is limited to
twice the number of metal wiring layers in the integrated circuit technology. The need for a
reasonable fabrication yield limits the number of metallization layers to $\sim 10$, which means
that the 2-D array can be at most 20 qubits wide. Fig. \ref{fig:qubitstripe} illustrates the case
for 5 metallization layers. In this respect, we agree with Copsey \textit{et al.} \cite{copsey03},
who pointed out this restriction specifically in the context of semiconductor qubits. Thus, while
the qubit array might be locally 2 dimensional, the overall architecture will consist of 1-D
stripes of moderate width, as illustrated in Fig. \ref{fig:serpentine}.

The lowest level of concatenated qubit encoding, $L=1$, can be laid out along stripe width, but all
higher concatenation levels must be laid out along the stripe length, and are effectively 1
dimensional. We are thus led to an essentially 1-D concatenation hierarchy, the most challenging
for internal quantum communication.

\begin{figure*}
\centering
\includegraphics[scale=0.75]{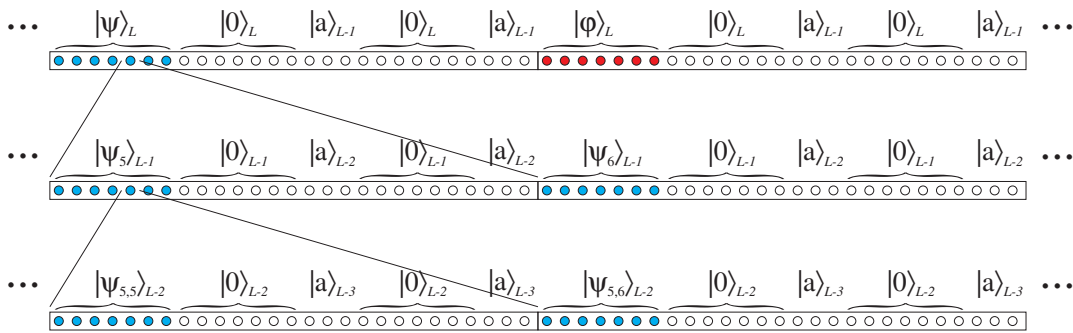}
\caption{\label{fig:linarray} A self-similar concatenated hierarchy
of logical qubits on a linear array, with concatenation level $L$
down to $L-2$ shown. Error correction requires a minimum of two
logical zeros, $|0 \rangle_L$, and six ancillae, $|a \rangle_{L-1}$.
Altogether, 27 level $L-1$ qubits are minimally required to protect
a single level $L$ qubit $|\psi\rangle_L$. The exponential growth
with concatenation level $L$ of \textit{physical} nearest-neighbour
operations to interact $|\psi\rangle_L$ and $|\phi\rangle_L$ is
apparent. We consider a layout with $L+1$ adjacent linear arrays of
qubits each organized according to the illustrated logical
heirarchy.}
\end{figure*}

Universal sets of fault-tolerant operations are known only for CSS
error correcting codes of various size
\cite{shor96,boykin99,aharanov99,steane03}. In our work, we shall
consider the $[7,1,3]$ CSS code. Concatenation \cite{knill96}, where
each logical qubit is composed of encoded qubits, which are in turn
composed of encoded qubits and so on, can suppress logical error
rate to arbitrary degree, provided the physical error rates remain
below a threshold value. The self-similarity of concatenation
naturally leads to the self-similar logical structure illustrated in
Fig. \ref{fig:linarray}. There are 7 level $L-1$ logical qubits
forming the CSS codeword that represents a single level $L$ logical
qubit $|\psi\rangle_L$. A minimum of two logical zeros,
$|0\rangle_L$, and six initially arbitrary ancillae,
$|a\rangle_{L-1}$, are required to perform error correction on
$|\psi\rangle_L$. We consider $L+1$ parallel lines of physical
qubits to implement error correction and computation with $L$ levels
of concatenation. The error correction protocol is described in
detail in the next section. An important feature of the self-similar
hierarchy is that at each concatenation level, the same qubit
protection block is employed (for ancillae as well as information
bearing qubits). Error correction can thus take place at any logical
level within an appropriate logical qubit protection block.

\section{Error Correction Protocol}

\begin{figure}
\centering
\includegraphics[scale=0.55]{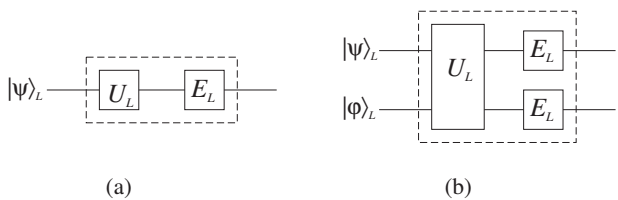}
\caption{\label{fig:circuiterror} Each unitary operation $U_L$ at logical level $L$ is followed by
error correction $E_L$ at error correction level $L$. }
\end{figure}

For estimating error thresholds, we consider an aggressive error correction scheme where every
unitary operation $U_L$ at concatenation level $L$ is followed by error correction $E_L$ at level
$L$, as illustrated in Fig. \ref{fig:circuiterror}.

\begin{figure*}
\centering
\includegraphics[scale=0.6]{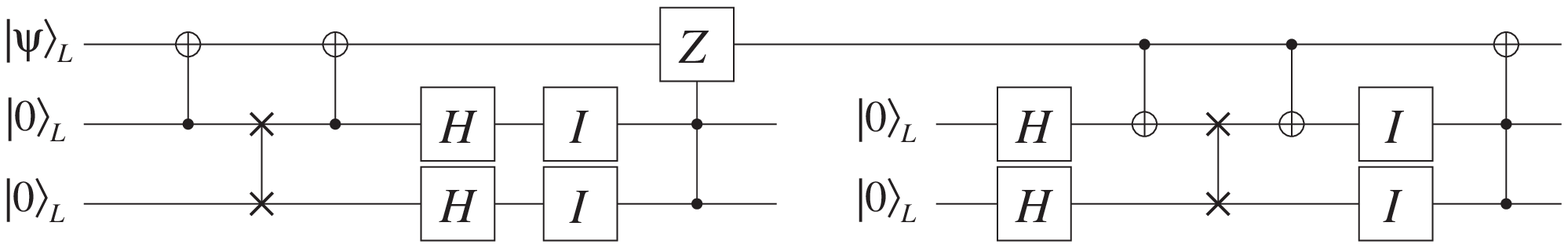}
\caption{\label{fig:EL} A modified Steane error correction circuit
($E_L$). The indicator block $I$ computes an error syndrome, and
decodes the syndrome into a bit-wise error indicator used for error
recovery. The logical SWAP gate, as well as the CNOT gates, requires
shuffling of the constituent $L-1$ qubits (see Fig.
\ref{fig:swaps}). We allow only nearest neighbour operations at all
logical levels in adherence to self-similarity. }
\end{figure*}

The error correction operation, $E_L$, can be implemented in a fault-tolerant manner with a Steane
error correction circuit \cite{steane97}, slightly modified to that shown in Fig. \ref{fig:EL}.
Error correction takes place within an error correction block, with the logical qubit
$|\psi\rangle_L$ and logical zero states $|0\rangle_L$ explicitly shown. The two groups of three
$L-1$ ancillae, $|a\rangle_{L-1}$, are made use of within the bit-flip \textit{indicator} circuit,
denoted by $I$. As can be seen in Fig. \ref{fig:EL}, the Steane error correction circuit is
particularly parsimonious in its use of gate operations, and leads to particularly favorable error
thresholds. The bit-flip indicator block $I$ is essential, where for each logical zero
$|0\rangle_L$ it computes a bit-flip error syndrome into three ancillae qubits $|a\rangle_{L-1}$.
The syndrome is then decoded within the indicator block $I$ into a bit-wise error indicator that
can be directly used for error recovery. Note also that only nearest-neighbour operations at logic
level $L$ are employed, in strict adherence to self-similarity from the physical layer up to
concatenation level $L$.

The key point about the bit-flip indicator block $I$ is that it operates on logical zeros that have
effectively measured the logical qubit error, but not the logical qubit itself, by virtue of a
logical CNOT gate. As was pointed out by Boykin \textit{et al.} \cite{boykin99b}, the
identification of which operations require full quantum coherence and which operations do not is
important since ``quantum'' operations require full protection against both phase-flip and bit-flip
errors, while ``classical'' operations require protection against bit-flip errors only. Note from
Fig. \ref{fig:EL} that the outputs of indicator block $I$ are used only as control bits for the
error recovery operations acting upon the logical qubit. Arbitrary phase flips in the output of $I$
have no effect on the logical qubit. Likewise, phase flips on the input of $I$ have no effect on
the logical qubit since the syndrome is encoded as bit-flips on the input to $I$. We need only
protect against bit-flip errors in $I$, so that the operations within $I$ can be thought of as
essentially ``classical'' in nature, even though they are executed by physical qubit gates. Thus,
$I$ can in principle be protected with classical fault tolerance, which has been shown to be much
more efficient than quantum fault tolerance \cite{boykin05}, to ensure that the operations within
$I$ will contribute negligibly to the quantum error threshold.

\begin{figure*}
\centering
\includegraphics[scale=0.6]{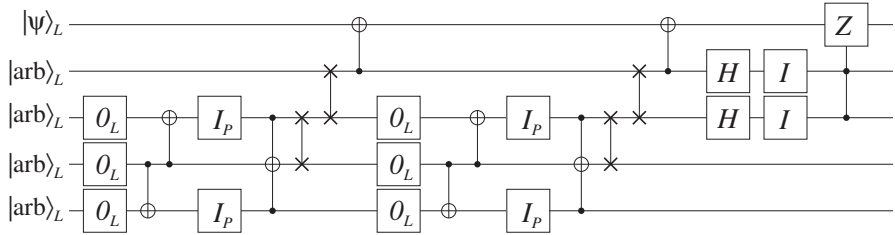}
\caption{\label{fig:fullEL} Error correction circuit (phase-error portion only) directly
incorporating the preparation of requisite logical zeros. Ancillae begin in arbitrary states
$|\mathrm{arb}\rangle$. Three $0_L$ blocks prepare logical zeros that are purified into a single
$|0\rangle_L$ state for use in error correction. A modified indicator block $I_P$ corrects for
possible parity errors in the raw $|0\rangle_L$'s. }
\end{figure*}

\begin{figure}
\centering
\includegraphics[scale=0.53]{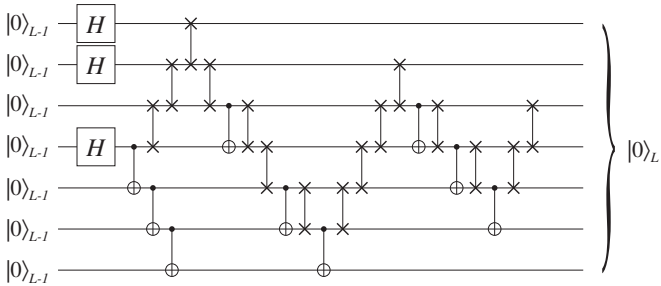}
\caption{\label{fig:logiczero} Circuit $0_L$ for preparation of a single logical zero $|0\rangle_L$
from lower level $|0\rangle_{L-1}$'s. Only nearest neighbour operations are employed. }
\end{figure}

Of course, the requisite logical zeros, $|0\rangle_L$, that allow for efficient fault-tolerant
error correction are complex entangled states which must be created with low error probability to
begin with. One approach to this problem is to dedicate adjacent quantum circuitry whose sole
function is to prepare and purify logical zeros, providing a steady supply at various concatenation
levels specifically for this purpose. Alternatively, the preparation of logical zeros can be
performed directly within the qubit error protection block. The full error correction circuit is
illustrated in Fig. \ref{fig:fullEL}. Purification of three $|0\rangle_L$'s, prepared by the $0_L$
block, results in a single $|0\rangle_L$ state for use in error correction. The $0_L$ zero
preparation block is given in Fig. \ref{fig:logiczero}. Bit-flip errors are corrected with a
modified indicator block $I_P$, which also corrects for a possible parity flip error corresponding
to the logical zero being in the state $|1\rangle_L$ (and thus requiring a minimum of 4 ancillae).
The qubit protection block must increase in size to accommodate $|0\rangle_L$ preparation in this
case. A total of 46 qubits would be required, arranged in the following sequence of $L-1$ qubits
(compare with Fig. \ref{fig:fullEL}): 7 qubits for storing $|\psi\rangle_L$, 7 qubits for storing a
$|0\rangle_L$, 3 ancillae $|a\rangle_{L-1}$ for $I$, 7 qubits for storing a $|0\rangle_L$, 4
ancillae $|a\rangle_{L-1}$ for $I_P$, 7 qubits for storing a $|0\rangle_L$, 7 qubits for storing a
$|0\rangle_L$ and 4 ancillae $|a\rangle_{L-1}$ for $I_P$.

\section{Error Threshold Penalty}

The number of physical qubits for our concatenated CSS encoding
required to store and protect one logical qubit is $27^L$ (or $46^L$
including logical zero preparation). Several levels of concatenation
already leads to a large number of physical qubits (although the
width of the qubit stripe grows only as $L+1$). Likewise, the number
of physical gate operations grows exponentially, $N^L$, where $N$ is
approximately the number of logical operations required at level
$L-1$ in order to implement a single logical function at level $L$.
For example, with a single level of encoding, $N$ is simply the
number of physical gate operations required to perform some function
on our 7-qubit CSS code word (or multiple code words in the case of
a multi-qubit logical function).

The number of gate operations $N$ will depend on the function being
performed. We consider implementing a simple two-qubit unitary,
$U_L$, followed by error correction, $E_L$, as illustrated in Fig.
\ref{fig:circuiterror}(b). Error correction might require $N=N_E$
logical gate operations at level $L-1$. There will be additional
logical SWAP operations at level $L-1$ required to move qubits
around since only nearest-neighbour interactions are permitted. We
let $N_{Ec}$ be the number of required nearest-neighbour SWAP
communication operations, bringing the total number of level $L-1$
operations to $N=N_E + N_{Ec}$. Of course, the unitary $U_L$ will
require $N_U$ operations at level $L-1$, as well as $N_{Uc}$
additional communication operations at level $L-1$. The total gate
operation count at level $L-1$ to implement $U_L$ followed by $E_L$
is simply $N=N_U + N_{Uc}+ N_E + N_{Ec}$. The total
\textit{physical} gate count is again approximately $N^L = ( N_U +
N_{Uc}+ N_E + N_{Ec} )^L$ because each of the $N$ operations at
$L-1$ is simply a unitary $U_{L-1}$ followed by error correction
$E_{L-1}$. The self-similar hierarchy requires that $N$ operations
at $L-2$ are required for each operation at $L-1$ and so forth,
including communication.

\begin{figure*}
\centering
\includegraphics[scale=0.65]{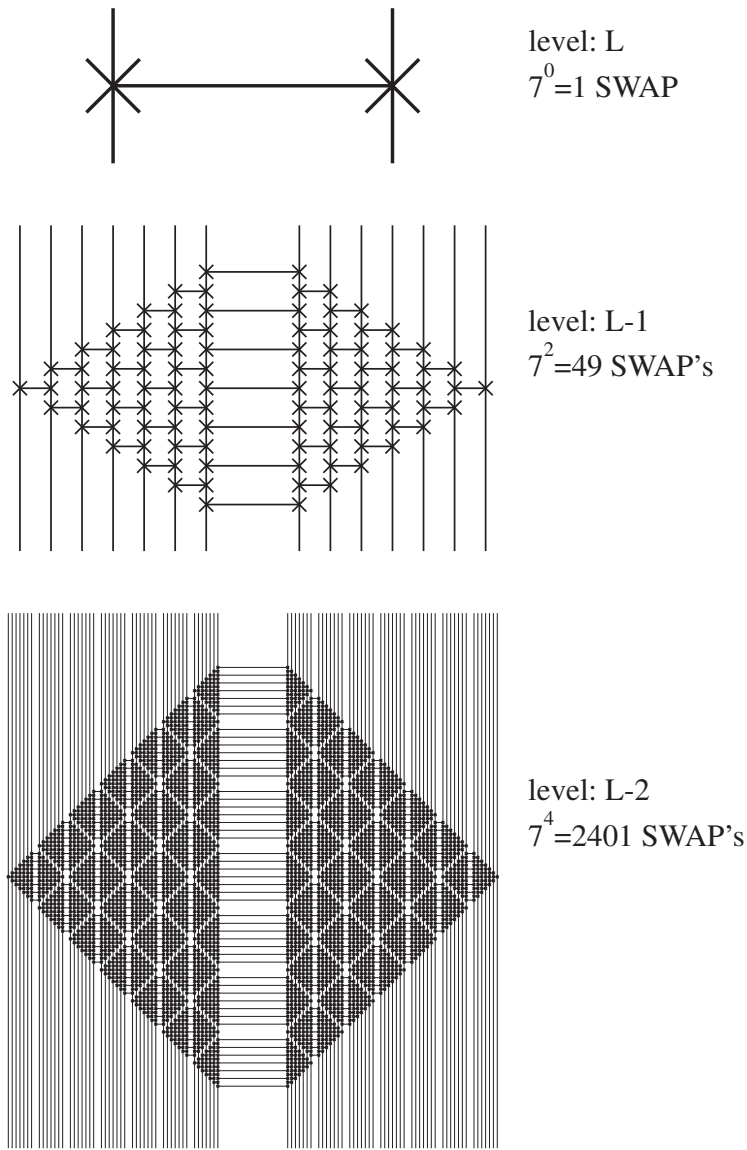}
\caption{\label{fig:swaps} A logical SWAP operation illustrated at concatenation levels $L$ through
$L-2$ with nearest neighbour interactions only. The number of level $L-1$ SWAPS required to
implement a single level $L$ SWAP between adjacent logical qubits is $N_U+N_{Uc}=7+42$. There are $21$ level $L-1$ SWAPs to
interleave the qubits, $7$ level $L-1$ qubit-wise SWAPs, and $21$ level $L-1$ SWAPs to undo the
interleaving. Note that a single gate failure does not produce correlated errors within a logical
qubit. Error correction, and swapping through the additional qubits in a qubit protection block, are
omitted here for clarity. }
\end{figure*}

\begin{figure*}
\centering
\includegraphics[scale=0.65]{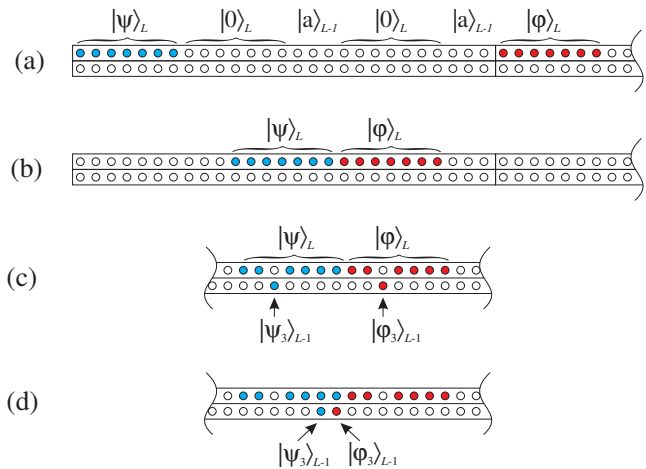}
\caption{\label{fig:cnotseq} Partial sequence for a logical level
$L$ CNOT operation illustrated at concatenation level $L-1$ with
nearest neighbour interactions only. The (a) logical code words
$|\psi\rangle_L$ and $|\varphi\rangle_L$ are (b) first brought into
adjacent positions, then (c) each of the 7 constituent $L-1$ qubits
are moved into an adjacent qubit row to be (d) brought together for
qubit wise interaction (only the third qubits $|\psi_3\rangle_{L-1}$
and $|\varphi_3\rangle_{L-1}$ are shown interacting). The logical
qubits are brought back to their original positions for error
correction after the logical CNOT. The scheme is applied recursively
until physical CNOT gates are performed in the $L+1^{st}$ row. The
CNOT gates for the error correction circuit are similarly
implemented. Note that a single gate failure does not produce
multiple errors within a logical qubit. }
\end{figure*}

In reality, the gate count $N_U+N_{Uc}$ varies among the various
logical qubit operations possible. For instance, Hadamard at level
$L$ requires $N_U=7$ Hadamard gates at level $L-1$ and $N_{Uc}=0$
communication gates. In contrast, the gate operations
$N_U+N_{Uc}=7+42$ involved in a logical SWAP on the same qubit line
are illustrated in Fig. \ref{fig:swaps} for adjacent logical qubits.
Clearly the number $N^L$ can be very large, although a substantial
fraction of operations at each logical level can be performed in
parallel. Note the fault tolerance of the logical SWAP gate: a
single swap gate failure induces one error in each logical qubit,
which can be recovered independently by error correction. Of course,
the extra qubits involved in a qubit protection block increases the
number of communication swaps $N_{Uc}$. As a final example, we show
the partial sequence of gate operations required for the logical
CNOT gate in Fig. \ref{fig:cnotseq}. It is in implementing the CNOT
gate that an additional line of qubits is used for every
concatenation level, resulting in a total of $L+1$ lines of qubits.
Similar sequences are used for the SWAP and CNOT gates required for
the error correction operation $E_L$, contributing to $N_E+N_{Ec}$.

\begin{table*}[t]
\centering
\caption{The gate count for error correction, $N_E+N_{Ec}$, and for logical CNOT operations, $N_U+N_{Uc}$, under different assumptions of internal communication resources and
quantum error correction. Approximate threshold gate error probabilities are given, as well as
control pulse accuracy thresholds (see text for details). \label{table:errors} }

\begin{tabular}{ll|c|c|c|c|}
& & & & & \\
& & Error & Two-Qubit & Error & Gate \\
& & Correction & Unitary & Probability & Accuracy \\
& & Gate Count & Gate Count & Threshold & Threshold \\
& & $N_E+N_{Ec}$ & $N_U+N_{Uc}$ & $P_{\mathrm{th}}=2/N^2$ & $\phi_{\mathrm{th}} = 2 \sqrt{P_{\mathrm{th}}} \times 180/\pi$ \\
& & & & &\\
\hline & & & & &\\
no communication ~ ~ & no $|0\rangle_L$ preparation & 70 & 7 & $3.4 \times 10^{-4}$ & $2.1^\circ$ \\
overhead & & & & &\\
& $|0\rangle_L$ preparation & 298 & 7 & $2.1 \times 10^{-5}$ & $0.52^\circ$\\
& & & & &\\
\hline & & & & &\\
remote CNOT & no $|0\rangle_L$ preparation & 238 & 35 & $2.7 \times 10^{-5}$ & $0.60^\circ$ \\
communication & & & & &\\
& $|0\rangle_L$ preparation & 1090 & 35 & $1.6 \times 10^{-6}$ & $0.14^\circ$ \\
& & & & &\\
\hline & & & & &\\
SWAP & no $|0\rangle_L$ preparation & 1008 & 203 & $1.4 \times 10^{-6}$ & $0.13^\circ$ \\
communication & & & & &\\
& $|0\rangle_L$ preparation & 3754 & 343 & $1.2 \times 10^{-7}$ & $0.034^\circ$ \\
& & & & &\\
\hline
\end{tabular}

\end{table*}

Despite the exponential increase in physical qubits and physical
gate operations with concatenation level (while the width of the
stripe merely grows linearly in concatenation level), logical errors
are suppressed double-exponentially with concatenation level. We let
$P_1$ be the logical error probability on a first level encoded
state, $|\psi\rangle_1$, after a two qubit unitary followed by a
single error correction cycle. By the fault tolerant construction of
$U_L$ and $E_L$, the probability of a logical error is bounded above
by the probability that two gate operations fail,
\begin{eqnarray}
P_1 \leq \left( \begin{array}{c}N\\2\end{array} \right) \epsilon^2 \simeq \frac{N^2}{2} \epsilon^2,
\end{eqnarray}
where $\epsilon$ is the probability of physical gate error, assumed
to be equal for all gates, and $N=N_U + N_{Uc}+ N_E + N_{Ec}$ as
before. While logical error rates shall vary slightly due to
differences in $N_U+N_{Uc}$ amongst the logical gate operations
with the dominant $N_{E}+N_{Ec}$ remaining fixed, a conservative estimate can be
had by taking the gate counts for the logical CNOT gate as
representative. The criterion for error correction to reduce the
likelihood of qubit error is $P_1<\epsilon$. This leads to the threshold error condition
$\epsilon<2/N^2$. Likewise, at higher levels of concatenation,
\begin{eqnarray}
P_L \leq \left( \begin{array}{c}N\\2\end{array} \right) P_{L-1}^2 \simeq \frac{N^2}{2} P_{L-1}^2,
\end{eqnarray}
leading to $P_{L-1} < 2/N^2=P_{\mathrm{th}}$ being the error threshold condition for all $L$. The
corresponding required phase accuracy for gate operations, as described in section
\ref{sec:gateaccuracy}, is $\phi =2\sqrt{2}/N$. From the above relations, we arrive at the standard
logical error probability for concatenated error correction,
\begin{eqnarray}
P_L \leq P_{\mathrm{th}} \left( \frac{\epsilon }{ P_{\mathrm{th}} } \right)^{2^L} \label{eqn:error}
\end{eqnarray}
but where $N$ now includes the nearest neighbour communication overhead at a particular
concatenation level. The exponent $2^L$ results in an overwhelming, \emph{super}-exponential in $L$
suppression of logical errors while the number of qubits and gate operations increase only
exponentially in $L$.

Suppose that a quantum computation requires a sequence of $T$ logical gate operations, then a
logical error probability $P_L = 1/T$ will give the correct result with only several trials of the
computation. The relation between the number, $T$, of operations in a calculation and concatenation
level $L$ can be written,
\begin{eqnarray}
T \geq \frac{ 1 }{ P_{\mathrm{th}} } \left( \frac{ P_{\mathrm{th}} }{ \epsilon } \right)^{2^L}
\end{eqnarray}
or alternatively,
\begin{eqnarray} L \leq \log_2 \left( \frac{\log_2(T
P_{\mathrm{th}})}{\log_2(P_{\mathrm{th}}/\epsilon)} \right)
\end{eqnarray}
For instance, the error threshold might be $P_{\mathrm{th}}=10^{-6}$
while the physical gate operation error is an order of magnitude
better, $\epsilon= P_{\mathrm{th}}/10 = 10^{-7}$. We then have an
accessible computation length $T=10^6\times10^{2^L}$, which for
$L=3$ gives $T\geq10^{14}$. It follows that interesting calculations
can be performed with only a few layers of concatenation (ie. a
qubit stripe with a width of only a few qubits) if physical error
probabilities well below the error threshold can be achieved.

The problem of estimating error threshold has been reduced to counting gate operations, for which
our numerical results are summarized in Table \ref{table:errors}. Note that we have neglected
storage errors in our present analysis since the coherence times of electron spins in
semiconductors \cite{tyryshkin03} exceed the expected gate operation times by at least $\sim 8$
orders of magnitude, with further improvement expected. The top row of Table \ref{table:errors}
gives the most favourable error thresholds where any qubit can interact with any other qubit
without any extra communication operations. The bottom row is the least favourable case where
nearest neighbour SWAP operations are used on a linear qubit array to implement all operations. The
middle row represents an intermediate case, where the remote-CNOT is used to perform a CNOT gate
between distant qubits \cite{sorenson98, gottesman98}. The remote-CNOT requires a shared EPR pair,
a resource that might be generated by independent hardware with sufficient purity that the EPR
error rate contributes negligibly to the overall error rate of the remote-CNOT and the error
threshold. Measurement and classical communication are also required for the remote-CNOT (see
appendix).

For all three communication schemes, the gate count is given in Table \ref{table:errors} for
sub-cases where $|0\rangle_L$'s are supplied by adjacent circuitry (a parallel qubit stripe, for
instance); or where the $|0\rangle_L$'s are prepared directly within the error correction circuit
itself (as in Fig. \ref{fig:fullEL}) thus burdening the error threshold. In the former case, we
assume that the adjacent circuitry can prepare and purify logical zeros to reach an error
probability much less than the preparation circuit of the former case, thereby contributing to the
error threshold negligibly. This might be achieved by successive rounds of purification.

In all cases, we assume that those portions of the circuit that can be implemented with classical
fault-tolerant logic \cite{boykin99b}, albeit with qubit gates, take advantage of the greater
efficiency of classical coding. The threshold error for classical fault-tolerant circuits has been
estimated to be between $\sim 1/100$ to $\sim 1/3000$ depending on topology and communication
resources \cite{boykin05}, we therefore assume the error rates in the classical circuits are
negligible compared to the quantum circuits, so that in counting the gate operations we can neglect
the operations in $I$ and $I_P$. Furthermore, the dual-control phase-flip ($\Lambda_2(Z)$) and
dual-control bit-flip ($\Lambda_2(X)=$Toffoli) are assumed to count merely as two-qubit
interactions, since fault-tolerant classical logic can be used to generate a single classical
control bit. The remaining sundry details involved in counting gate operations are left to the
appendix.

Observing the gate error thresholds in Table \ref{table:errors}, we see that SWAP communication
incurs a penalty of $\sim 175$X compared to the case of free communication. Communication through
the remote-CNOT incurs a penalty of $\sim 12$X compared to the free communication case. The
improvement associated with remote-CNOT communication is not as much as one might expect, since the
remote-CNOT requires multiple operations proportional to the size of the logical qubits. Thus,
internal quantum communication reduces gate error thresholds for fault tolerant computation by a
substantial factor that we estimate to be from $\sim 12$X to $\sim 175$X. While this certainly
increases the difficulty in experimentally realizing fault tolerant gate operations, it is by no
means an impasse for solid state quantum computation, as we discuss in the next section.

\section{Error Probability and Gate Operation Accuracy}
\label{sec:gateaccuracy}

\begin{figure}
\centering
\includegraphics[scale=0.55]{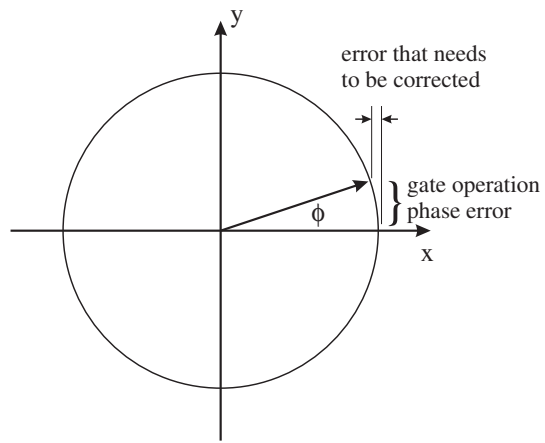}
\caption{\label{fig:spin} A conceptual illustration of a qubit pseudo-spin that might miss a target
x-axis by an angle $\phi$ due to a control pulse error. The resulting probability of qubit error is
$\epsilon \approx (\phi/2)^2$. }
\end{figure}

So far, we have worked entirely with error probabilities. In practice, experimental gate accuracy
is more naturally specified in terms of control pulse amplitude. Consider the spin (or a qubit
pseudo-spin), illustrated in Fig. \ref{fig:spin}. Suppose a control pulse, as used in spin
resonance, was to bring the spin into alignment with the x-axis. However, an error in pulse area,
phase, or timing may cause a misalignment by some small angle $\phi$. The probability of error,
$\epsilon$, is then the probability that the spin is not projected into the +x direction when a
measurement is performed along the x-axis. The probability of projection along the +x direction is
$\cos^2(\phi/2)$, so that the error probability is,
\begin{eqnarray}
\epsilon = \sin^2(\phi/2) \approx (\phi/2)^2.
\end{eqnarray}
The required gate timing and amplitude accuracy is $\phi = 2\sqrt{\epsilon}$, specified as a phase
angle, is proportional to the \emph{square root} of the threshold error probability. The gate
accuracy thresholds are given in degrees in Table \ref{table:errors}. Of course, the $\sim 12$X to
$\sim 175$X penalty in error probability threshold becomes only a $\sim 3.5$X to $\sim 13$X penalty
in control pulse accuracy. In order to achieve an error probability of $10^{-7}$, one would require
about 1/30 of a degree accuracy in control pulse timing, which is not entirely infeasible since it
would require about 1 picosecond phase accuracy in a clock period of about 10 nanoseconds. Recall
that an error probability of $10^{-7}$ for a quantum processor with threshold error probability
$10^{-6}$ and 3 levels of concatenation will allow a computation with $\geq 10^{14}$ operations.
Thus, thinking about gate errors in terms of phase angle makes it clear that very small error
probabilities are achievable.

\section{Conclusions}

Internal quantum communication remains a challenging architectural problem that impacts the
threshold error for fault-tolerant computation with encoded logical qubits. The communication
operation overhead required to distribute information among a number of qubits that grows
exponentially with concatenation level can be a significant burden. Whether one is limited to
nearest-neighbour communication, a communication bus (as in the original Cirac-Zoller ion trap
proposal \cite{cirac95}), or communication by modified teleportation schemes such as the
remote-CNOT, there is always a communication penalty in error threshold. The minimum communication
overhead cost is associated with a communication bus, where a single operation for ``transmitting''
and a single operation for ``receiving'' is possible in principle. The question of whether a
sufficiently robust communication bus is available for solid state qubits remains open. Ballistic
transport of electron spins through mesoscopic wires is predicted to give error rates of $\sim 0.6$
for GaAs \cite{recher00}, far above our stated threshold requirements even for the free
communication case. Much more promising is the combination of cavity QED techniques with confined
electron spins \cite{imamoglu99} or superconducting circuits \cite{wallraff04}, where an
electromagnetic bus can couple a number of qubits. The error rates of such a bus, the
reconfigurability of its links, and its parallelism (ie. how many qubits can be transported
simultaneously? through the same link?) must all be carefully considered in determining what
benefits, if any, we can expect over nearest neighbour architectures. Nonetheless, we expect that
communication overhead can be mitigated to a large extent by circuit optimization. Recent work
\cite{fowler04} on laying out Shor's factorization algorithm on a linear chain of qubits under the
restriction of nearest neighbour interaction has shown that circuit optimization can greatly reduce
the number of logical qubit SWAPs required.

\section{Appendix - Threshold Error Calculations}

We provide a brief summary here of the counting of gate operations, which then leads to the
threshold error. Error correction at concatenation level $L$ with the circuit $E_L$ requires the
use of both single qubit unitaries and two qubit unitaries at levels $L$ down to the physical
layer. Interestingly, the quantum portions of the circuit $E_L$ (see Figs. \ref{fig:EL} or
\ref{fig:fullEL}) consists of gate operations that are directly fault tolerant, where qubit-wise
(or transversal) operations are sufficient. These operations include CNOT, SWAP, and H (Hadamard
rotation). The control bits of the dual control gates are classical, so a full quantum Toffoli is
not required. Of course, indirectly fault tolerant gates such as the Toffoli ($\Lambda_2(X)$) or
$\pi/8$ rotation $(Z^{1/4})$ are required for universal computation. We do not calculate the error
threshold for indirectly fault tolerant gates here.

\subsection{Free Communication}

First, we consider the idealized case where communication is achieved without any extra operations,
in other words, any two-qubits can interact directly at any time. In this case, $N_{Uc}=N_{Ec}=0$
and we need only count the number of computationally useful gates. A directly fault tolerant
two-qubit unitary will require $N_{U}=7$ operations. The error correction gate count without
logical zero preparation is,
\begin{eqnarray}
N_E = 4 \times 7 \mathrm{CNOT} + 4 \times 7 \mathrm{H} + 7 \Lambda_2(X) +7 \Lambda_2(Z)=70 \nonumber \\
\end{eqnarray}
where the $L-1$ gate type and count are indicated. With logical zero preparation, we have,
\begin{eqnarray}
N_E &=& 70 + 12 \times 0_L + 4 \times 7  \Lambda_2(X) + 8 \times 7 \mathrm{CNOT} \nonumber \\
&=& 70 + 12\times(3 \mathrm{H} + 9  \mathrm{CNOT}) + 84 \nonumber \\
&=& 298
\end{eqnarray}
where again $L-1$ gate type and count was indicated.

\subsection{remote-CNOT communication}

\begin{figure}
\centering
\includegraphics[scale=0.55]{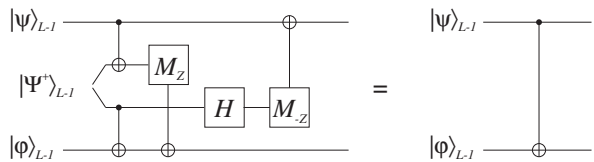}
\caption{\label{fig:recnot} The remote CNOT gate, modified from \cite{gottesman98}, requires a
shared EPR pair, $|\Psi^+\rangle=(|01\rangle + |10\rangle)/\sqrt{2}$, measurement, $M_Z$, and
classical communication to implement a CNOT operation between distant qubits.}
\end{figure}

Next, we consider the intermediate communication case involving remote-CNOT operation, which we
abbreviate as reCNOT. The reCNOT circuit is indicated in Fig. \ref{fig:recnot}. For simplicity, we
assume that the classical communication and EPR preparation introduce negligible errors compared to
the other gate operations involved. We see that a reCNOT between two level $L-1$ qubits requires 5
level $L-1$ operations, so that a reCNOT between two level $L$ qubits requires
$N_U+N_{Uc}=5\times7$ level $L-1$ operations. The error correction gate count without logical zero
preparation becomes,
\begin{eqnarray}
N_E &=& 4 \times 7 \mathrm{reCNOT} + 4 \times 7 \mathrm{H} + 7 \Lambda_2(X) +7 \Lambda_2(Z) \nonumber \\
&=& 140 + 28 + 35 + 35 \nonumber \\
&=& 238
\end{eqnarray}
where $\Lambda_2(X)$ and $\Lambda_2(Z)$ are counted as reCNOT operations (recall they can be
implemented with single classical control bits). With logical zero preparation, we have,
\begin{eqnarray}
N_E &=& 238 + 12 \times 0_L + 4 \times 7 \Lambda_2(X) + 8 \times 7 \mathrm{reCNOT} \nonumber \\
&=& 238 + 12\times(3 \mathrm{H} + 3 \mathrm{CNOT} + 6 \mathrm{reCNOT}) \nonumber \\ &&+ 140 + 280 \nonumber \\
&=& 238 + 432 + 140 + 280 = 1090
\end{eqnarray}
where we have made use of both nearest neighbour CNOT and reCNOT in the logical zero preparation.

\subsection{SWAP communication}

Finally, we consider communication by SWAP gates. Without logical zero preparation, a level $L$
qubit protection block is 27 $L-1$ qubits long. Applying CNOT between two level $L$ qubits as in Fig. \ref{fig:cnotseq} requires
$N_U+N_{Uc}=203$ level $L-1$ operations on each logical qubit argument. The error correction operation requires,
\begin{eqnarray}
N_E &=& 4 \times (7\mathrm{CNOT}+112\mathrm{SWAP}) + \nonumber \\
&& 4 \times 7 \mathrm{H} + 2\times(7 \mathrm{SWAP}+84\mathrm{SWAP}) + \nonumber \\
&& (7 \Lambda_2(X)+154\mathrm{SWAP}) +(7 \Lambda_2(Z)+154\mathrm{SWAP}) \nonumber \\
&=& 1008
\end{eqnarray}
where we note that $112$ communication SWAPs are required for applying CNOT between $|\psi\rangle_L$ with an
adjacent $|0\rangle_L$, and $84$ communication SWAPs are required for logical swapping of a $|0\rangle_L$
with another $|0\rangle_L$ taking account of the extra ancillae $|a\rangle_{L-1}$ in the way.

When logical zero generation is included, the qubit protection block increases in size to 46
qubits. Applying CNOT between two level $L$ qubits now requires $N_U+N_{Uc}=343$ level $L-1$
operations because of the increased size of the qubit protection block. The error correction
operation requires,
\begin{eqnarray}
N_E &=& 1008 + 12 \times 0_L + 2 \times (7\mathrm{SWAP}+84\mathrm{SWAP}) + \nonumber \\
&&4 \times (7\mathrm{SWAP}+98\mathrm{SWAP}) + 4 \times (7\mathrm{CNOT}+ \nonumber \\
&& 112\mathrm{SWAP}) + 4\times(7 \mathrm{CNOT}+168\mathrm{SWAP}) +  \nonumber \\
&& 4\times(7 \Lambda_2(X)+154\mathrm{SWAP}) \nonumber \\
&=& 3754
\end{eqnarray}
where we note that each logical $|0\rangle_L$ generation requires 27
level $L-1$ operations (Fig. \ref{fig:logiczero}), and the SWAP communication accounts for all
extra ancillae $|a\rangle_{L-1}$ in the way.

\section*{Acknowledgment}

We thank Isaac Chuang for bringing the issue of internal
communication in quantum computation to our attention. We also thank
Daniel Gottesman for pointing out an error in an earlier manuscript.
This work was supported by the Defense Advanced Research Projects
Agency and the Defense MicroElectronics Activity.

\bibliographystyle{IEEEtran}

\end{document}